\documentclass{article}
\usepackage{graphicx}
\usepackage{epsf}
\usepackage{amsfonts}
\usepackage{times}

\title{Scrambling Time and Causal Structure of the Photon 
Sphere of a Schwarzschild Black Hole}
\author{Peter W. Shor\\
MIT\\
Cambridge, MA 02482}

\begin{document}
\maketitle
\vspace{1in}

\noindent
{\bf Abstract:} 

%Check: black holes as Gaussian bosonic channel.

Recently, physicists have started applying quantum information theory to black holes.
This led to the conjecture that black holes are the fastest scramblers
of information, and that they
scramble it in time order $M \log M$, where $M$ is the mass of the black
hole in natural units. As stated above, the conjecture is not completely defined,
as there are several possible definitions of scrambling times. It appears that not all
papers that refer to this conjecture interpret it the same way. We consider a definition
of scrambling time stronger than the one given in the paper that first proposed this 
conjecture [Sekino and Susskind, {\em J. High Energy Phys.}  {\bf 0810}:065 (2008)],
and show that this stronger version of the conjecture appears to be incompatible with
a number of other widely-believed and reasonable-sounding properties of black holes.

We argue that for the scrambling time of a black hole to be this
fast, either relativity is violated or
non-standard physics must be occurring outside the stretched event horizon of a
black hole. More specifically, either information is being transferred
faster than relativity would permit, the information is not carried by
the Hawking radiation and thus must be carried by
unknown physics, or the Hawking radiation carries much more information
than standard thermodynamics would permit. 

We analyze the situation from the viewpoint of an outside observer who never
falls into the black hole. We assume that an outside observer
never sees anything actually fall into the black hole. We also assume that
from the viewpoint of this observer, the physics near a black hole are very much 
like the known laws of physics, except possibly in the stretched horizon, where physics
at Planck-scale energies starts being relevant. For the physics near the
horizon, all we assume is that the structure of space-time is roughly that
predicted by general relativity.

Unlike Suskind's complementarity principle, we
do not require that an outside observer's point of view be in any way
compatible with that of an observer falling into the black hole. In fact,
we do not consider the viewpoint of observers falling into the black hole
at all. Thus, some of the solutions that have been proposed to evade the contradiction
discovered in the AMPS paper [Almheiri et al., {\em J. High Energy Phys.}  
2013:62 (2013)], in particular the assumption
that black holes have firewalls just below
the horizon that destroy any infalling information do not appear to
address the problems our paper raises. 

Naturally, in order to show this, we need to make some assumptions.
Our first assumption is that, outside the stretched horizon, the laws of
physics are well approximated by some quantum field theory. The second is
that the causality structure of spacetime outside the horizon is dictated by
the laws of general relativity. Third, we
assume that the Hawking
radiation carries the information that exits the black hole, as well as the
information involved in scrambling the black hole. While we allow
information to be stored by a different 
means in the stretched horizon, general relativity does not appear
to permit fast scrambling
unless this information leaves the stretched horizon.  Finally, we assume 
that the usual laws of thermodynamics govern how much information can 
be carried by  Hawking radiation.

%Let us point out that several definitions of scrambling time have been
%studied. The argument in our paper only applies to the scrambling time
%as defined by entanglement, and not as defined by
%out of time order correlations. While it seems to be widely believed that these
%two definitions of scrambling time give approximately the same time, we do
%not believe that this is the case.

Our argument considers the structure of the photon
sphere from the point of view of an outsider who stays outside the black
hole.
The basic idea of our argument is to divide the photon sphere into cells,
and use computer-science style arguments to show that it will take at least
order $M^2$ time to transmit enough information from one side of the black
hole to the other so as to maximally entangle the two sides.

\section{Introduction}

The black hole information paradox is the question of whether the dynamics
of black hole evaporation is unitary, or whether information is lost when
you throw it into a black hole. Classical general relativity says that
anything that is thrown into a black hole can never come out, while
quantum mechanics says that physical processes are reversible, so that
information is never destroyed. In \cite{Bekenstein}, Bekenstein 
proposed that black holes obey their own laws of thermodynamics, and that
the entropy of a black hole was of order $M^2$ bits, where $M$ is the mass
of the black hole. In
\cite{Hawking}, Hawking continued the investigation of the thermodynamics
of black holes and argued that  
radiation emerges from a black hole, and thus that the black hole will
eventually evaporate after time order $M^3$. While
Hawking's argument does not appear to let information escape 
from a black hole, his argument is semi-classical, and thus can only be
an approximation to the true physics. It is thus not clear 
at present whether information can escape from
a black hole in a full theory of quantum gravity. Hawking radiation is very
similar to Unruh radiation, the radiation that an accelerating observer in 
a vacuum state sees. While some treatments of black holes distinguish
between Hawking and Unruh radiation, classifying the photons
that escape the black hole as ``Hawking radiation'' and the virtual photons
that remain inside the black hole as ``Unruh radiation'', we will not. We use
the term ``Hawking radiation'' for both of these phenomena when they 
are in the vicinity of black holes. 

Maldacena \cite{Maldacena} discovered a correspondence between
Anti de Sitter theories with gravity and conformal field theories without
gravity (called AdS-CFT). Since conformal field theories are unitary, this 
implied that theories of quantum gravity should also be unitary.

In order to reconcile general relativity and quantum mechanics,
Susskind \cite{complementarity} proposed a complementarity
principle, where both an observer remaining outside the black hole and an
observer falling in see things happening in accordance with the laws 
of physics, but where these two observers do not necessarily agree 
on exactly what happens. This complementarity principle has been seriously challened
by an argument put forth in 2013 in a paper generally known as AMPS (from its authors'
initials) \cite{AMPS}. More specifically, they use the monogamy of
entanglement \cite{monogamy} to argue that the information inside a black hole cannot
be entangled both with the information near the horizon and with the
earlier Hawking radiation. They then argue that this means that information 
cannot escape from a black hole without producing a ``firewall'' across the
horizon, where any infalling matter is destroyed.
% [Mention ``strong complementarity'' here. \cite{BF}]

Reasoning about black hole dynamics and complementarity,
in 2007 Hayden and Preskill \cite{HP}  gave arguments for why the scrambling time
of a black hole had to be at least order $M \log M$, where $M$ is the
mass of the black hole in natural units.\footnote{Natural units are chosen so that 
$G = c = \hbar = k_B = 1$, where $G$ is the gravitational constant, $c$ is
the speed of light, $\hbar$ is Planck's constant, and $k_B$ is Boltzmann's
constant. All quantities in this paper will be given in natural units.}  
An informal definition of
the scrambling time is how
fast the information in a black hole gets ``mixed up''. This paper led to a
more serious study of scrambling time. 
Several definitions of scrambling time have been proposed. These
do not necessarily all yield the same quantity for the scrambling time \cite{LLZZ}. 
These definitions will be discussed later in the paper.

Hayden and Preskill were reasoning about whether it was possible
to use black holes to observe a violation of the no-cloning theorem.
They assume that an observer, Bob, knows the exact state of a black hole. 
He then throws a quantum state into the black hole.
Hayden and Preskill showed that if he waits for the scrambling time,
Bob can use the Hawking radiation the black hole emits
to recover the information he threw in. Bob finally jumps into the 
black hole in an attempt to catch up with the information he earlier
threw in.  If he can successfully do this, then he would have two copies
of the quantum state, and thus would have
effectively cloned a quantum state, a violation of the unitary principle
of quantum mechanics (although nobody outside the black hole could 
verify that he has been successful at this).

What Hayden and Preskill showed was that as long as Bob has to wait
a time of at least order $M \log M$, he can never catch up with the
information thrown into the black hole. The fact that 
Hayden and Preskill needed about
the scrambling time was thus that it was at least order $M \log M$, as this
corresponded to the worst case of their analysis, where Bob comes the
closest to catching up with the information he threw in. 

A paper
of Seskino and Susskind \cite{SS} followed up on the Hayden and Preskill 
paper. They gave a definition of scrambling time that seemed to be
the minimum needed for the Hayden-Preskill argument to work,
and made the conjecture that black holes were indeed fast
scramblers, and mixed information up in order $M \log M$ time. Some 
support for this idea was presented in \cite{butterfly}. 

One argument given in \cite{HP,SS} for why the scrambling
time should be order $M \log M$ was that if you add mass or
electric charge to the black hole, it equalizes in a time of order $M \log M$.
Thus, the time for information scrambling should also be $M \log M$.
We believe this is a misleading argument. 
To see why, consider the analogy of putting a drop of dye into a pitcher of
water.  The water level will equalize in a matter of seconds, while it takes
much longer for the dye to diffuse evenly throughout the water---certainly 
at least several minutes. 
This is because the process of equalizing the water level is driven by an
energy difference, while the diffusion of the dye does not change the
energy of the system. Similarly, distributing
mass or electric charge uniformly around the black
hole decreases its energy, while there is no energy decrease associated
with spreading information around the black hole. 

Another argument for why the scrambling time should be order
$M \log M$ arises from the AdS-CFT correspondence. Shenkar and Stanford
\cite{butterfly} compute time-ordered correlation functions in the CFT side
of the correspondence, and show that the time scale of the decay of
these correlations is order $M \log M$. They conclude that the scrambling time
on the CFT side of the correspondence is order $M \log M$, and thus, that
the scrambling time
in the AdS side of the correspondence should also be order $M \log M$. 
While we agree that it seems  that the out-of-time-order correlations probably do
decay with time scale order $M \log M$, we do not see why the timescale for the decay of
time-ordered correlation functions should be the same as the scrambling
time, especially for the stronger definitions of scrambling time that have been
proposed.

In this
paper, we try to assume that physics near a black hole
behaves as much as possible like
established physics. We assume
that the black hole dynamics are those seen by an outside observer, and that
nothing actually ever reaches the event horizon. We assume that the
structure of the space-time outside the black hole is well-approximated by
the laws of general relativity. We assume further
that the black hole dynamics is unitary; that outside the stretched horizon,
all information
the black hole is carried by the Hawking radiation; and that the
thermodynamics of Hawking radiation is given by the standard formulas for
the thermodynamics of thermal radiation. 
Other than the laws of general relativity, we try to make no assumption on 
the physics in the stretched horizon, where the energy scales are the Planck
energy.

Using a definition of scrambling time stronger than Sekino and Susskind,
but comparable to definitions considered in several other papers that discuss
the scrambling conjecture,
we show that under
these assumptions, for black holes to scramble information faster than
order $M^2$,
they must be able to transmit information from one part of the black
hole to another faster than the speed of light, a 
violation of causality in relativity theory. 

%There are several possible ways to define scrambling times. One is to
%look at two-point, or $k$-point correlation functions, and see how quicklySchwarzschild
%they go to $0$. Another is to measure the time it takes the state to become
%distributed nearly uniformly in the Haar measure. A third is to wait until
%the density matrix of the average state is close to the identity matrix. In
%general, these definitions do not give the same scrambling time \cite{LLZZ}. 
%We will use the time it takes for a state with two nearly untangled
%halves to reach nearly full entanglement. This is the definition used by
%Seskino and Susskind. \cite{SS} 

Let us note that, under the assumptions of relativity, information cannot be
transmitted quickly in the stretched event horizon of the black hole. Suppose
we constrain the information to move around the black hole at a height $h$ or less above
the horizon, i.e., at radius at most $2M+h$. The line element for Schwartschild coordinates is
\[
ds^2 = -\left(1-\frac{2M}{r}\right)dt^2 + \left(1-\frac{2M}{r}\right)^{-1} dr^2
+r^2  (d\theta^2 + \sin^2 \theta d\phi^2). 
\]
If we set $ds^2 \leq 0$, to specify a lightlike or timelike trajectory,  then we see
that 
\[
r^2( d\theta^2 + \sin^2 \theta d\phi^2)\leq \left(\frac{h}{2M+h}\right) dt^2,
\]
showing that in time $t$, we can move a distance of at most $t\sqrt{ h/(2 M)}$ radially.
The stretched horizon is of order the Planck distance above the event horizon, which
corresponds to $h = 1/M$. Thus, to transmit information that stays within the
stretched horizon from one side of a black hole to another, relativity says that we
need time order $M^2$. One of the assumptions we make in this paper is that
this bound holds. 

\section{Scrambling Time}
Let us consider two hemispheres
of the black hole, which we will call the north and
south hemisphere. In a Haar-random pure state, these two hemispheres
will have nearly maximal entanglement; that is, since a black hole has
order $M^2$ bits of entropy, they will have order $M^2$ bits of entanglement.  
It seems likely that there is some minimum
entanglement between these two hemispheres in all low-energy states,
but this should be governed by the boundary, and thus
be on the order of $M$ bits, much less than the maximal
entanglement.  Assume that the two hemispheres start out in a random
nearly unentangled state. We ask how long it will take the state to evolve 
to a nearly maximally entangled state. This is our definition of scrambling time.

One might object that it is hard to initialize the black hole in a nearly 
unentangled state, i.e., a state having much less than $M^2$ 
entanglement. To address this objection, let us note that if such
states exist, one can ask the question of how long it will take these states
to evolve into nearly maximally entangled states; one does not need
a plausible physical mechanism for producing these states. We
see no reason why these states should not exist.

There are many other possible definitions of scrambling time. One is simply
the time scale it takes for out-of-time-order correlations to decay. This is the definition
used in \cite{butterfly,SBSSH}.
Another is the time it takes to reach a maximally entangled state, starting not
with a product state of two nearly equal halves, but with a tensor product 
of a pure state on one qubit with a Harr-random pure state on the remaining qubits. 
This last definition is close to Sekino and Susskind's \cite{SS}.
We do not see any reason why the scrambling time should not be of order $M \log M$
for these definitions. Sekino and Susskind chose this definition so that it would
be close to what would 
is needed to be able to recover the information in the Gedankenexperiment
of Hayden and Preskill \cite{HP}; however, we don't know of any proofs
that this definition of scrambling is sufficient to let one recover the information
in order $M \log M$ time. The paper \cite{LSHOH} defines the scrambling time in a way much
closer to the one we use in this paper: the time it takes to evolve from a tensor product
of $n$ qubits each in a pure state to a state
that is nearly maximally entangled on subsystems of size $\kappa n$ for some constant
$\kappa$. 
The paper \cite{YK} assumes that the quantum system is in a 
Haar-random state after the scrambling time, also a stronger criterion than our
definition.

\section{The Cell Structure}

The {\em photon sphere} of a Schwarzschild black hole 
is defined as everything inside the smallest possible circular orbit but
inside the horizon. For a
Schwarzschild black hole, which has radius $2M$, the smallest possible 
circular orbit has radius $3M$.
What we do is first 
divide the photon sphere the black hole into cells, with the property
that, as seen from an observer far from the black hole, information can be transmitted
from one part of a cell to any other part of the same cell in time order $M$. 
We then calculate that the Hawking radiation within each cell only
can carry a constant number of bits. Thus, we can 
model the causality structure of the black hole as a distributed
network of processors, where each processor only contains a constant
number of bits, and takes time order $M$ (as measured by an 
observer far from the black hole) to communicate with its neighbors.

We show that this distributed network cannot transmit 
information from one side of the black hole to the other
quickly enough to be a fast scrambler. 

In order to give bounds on the flow of
information in the photon sphere, we divide the photon sphere into cells.
The cells (from the point of view of an accelerating observer hovering within
a cell) should be roughly constant diameter in all directions. And from
the point of view of an outside observer far from the black hole, 
each round trip from a face 
of the cell to the opposite face and back should take time order $M$. 

\begin{figure}[tbhp]
\vspace{-.2in}
\begin{center}
\includegraphics[width=2.1in]{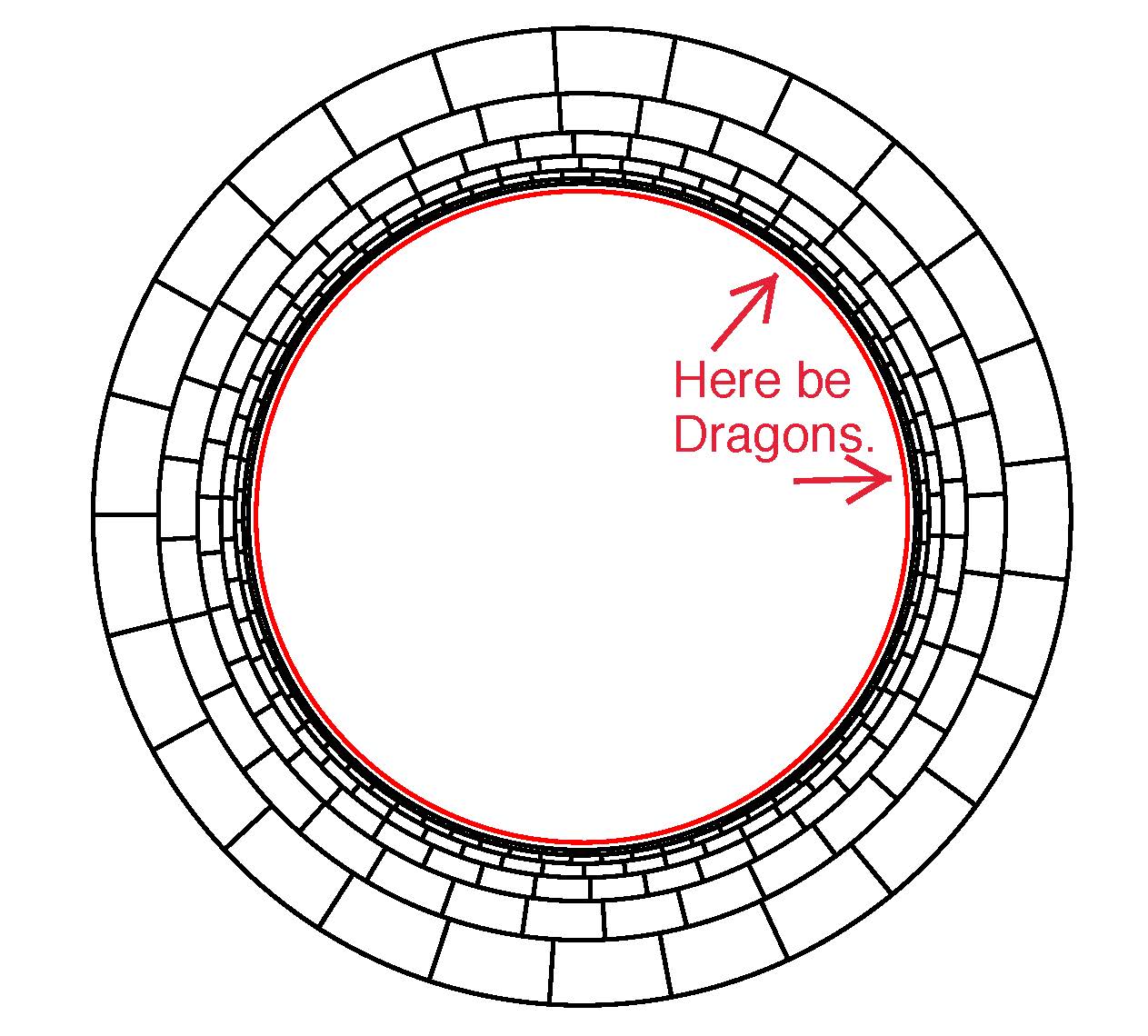}
\end{center}
\caption{The black hole cell structure in Schwarzschild coordinates, 
depicted in two dimensions. Note that
the aspect ratio of the cells appears larger as you approach the horizon.
This is an artifact of the Schwarzschild coordinates; while the cells do
grow smaller as you approach the horizon, an observer hovering 
near the horizon would see these cells as having an aspect ratio of
near unity.
}
\label{figure-cells}
\end{figure}

Recall that the line element of the Schwarzschild coordinates is
\[
ds^2 = -\left(1-\frac{2M}{r}\right)dt^2 + \left(1-\frac{2M}{r}\right)^{-1} dr^2
+r^2  (d\theta^2 + \sin^2 \theta d\phi^2). 
\]
Let $h$ be the height above the horizon, i.e. $h = r-2M$. For an
observer who stays at the same $r, \Omega$ in Schwarzschild coordinates,
we have
\[
d\tau = \sqrt{1-\frac{2M}{r}} dt = \sqrt{\frac{h}{r}} dt.
\]
In the photon sphere, $2M \leq r \leq 3M$, so 
\[
\sqrt{\frac{h}{r}} \approx \sqrt{\frac{h}{M}}.
\]
Thus, the diameter of these cells in Schwarzschild coordinates
in the direction parallel to the horizon 
should be order $M \sqrt{h/M}$. 
Because radial distance is lengthened 
by an additional factor of order $\sqrt{h/M}$, 
the vertical diameter of these cells
in Schwarzschild coordinates order is $M (h/M)$. 
Thus, the cells have diameter order $\sqrt{Mh}$ in the
direction parallel to the horizon, and diameter order $h$ in
the vertical direction.
For a two-dimensional representation
of what these cells look like, see Figure \ref{figure-cells}. 

It follows that 
the number of cells that cross a great circle
at radius $2M+h$ is of order 
\[
\frac{2M+h}{\sqrt{Mh}} \approx \sqrt{M/h},
\]
and the number of cells that cross a line from the photon sphere 
to a point at
$2M+h$ is order
\[ \log({M/h}).
\]

Let us now calculate the entropy, i.e., the number of bits contained
in the Hawking
radiation in each cell. The entropy of black-body radiation in a cell of
volume $V$ is 
\[
S = \frac{4 \pi^2}{45} VT^3,
\]
and so is proportional to $VT^3$, where $T$ is the temperature. The
side length of each of these cells, as seen from a stationary observer,
is of order $\sqrt{M h}$, so the volume is
of order $M^{3/2} h^{3/2}$. 

What is the temperature? The temperature seen by a
near-horizon observer (which still holds within a constant factor
for $h < M$) is 
\[
T \approx \frac{1}{4 \pi \sqrt{2 M h}}
\]
Thus, the number of bits in each cell is proportional to
\[
V T^3 =O(1)
\]
and so approximately constant. This radiation must consist of virtual particles,
since it does not contribute to the mass of the black hole. However, virtual
particles can have observable effects, and thus presumably can carry 
information.

Conventional wisdom says that the continuous nature of space and time breaks
down at the Planck scale. The value of $h$ which gives Planck-scale cells
is order $1/M$. If we stop the process at $h=1/M$, there are order $M^2$ cells, a
constant fraction of them just above the horizon of the black hole.
The exact constant factor on the $M^2$ depends on exactly
where we stop dividing the
cells.

Black hole thermodynamics predicts that the entropy of a black hole is
$A/4$, where $A$ is the surface area.  So if we
assume that each cell contains a constant number of bits of Hawking radiation, then 
for an outside observer, the entropy encoded in the Hawking radiation is 
sufficient to account for the black hole's entropy. This explanation for the entropy has
been previously proposed \cite{Jacobson, Wald}.  Is the entropy really
encoded in the Hawking radiation this way?
While we may have to wait until there is
a microscopic theory of quantum gravity to answer this question
definitively, it seems consistent with our current knowledge of physics.

Recall Landauer's precept ``information is physical.'' In order for the black
hole's quantum state to be scrambled, information must be carried from one
part of the black hole to another. And if we accept Landauer's precept, it
must be carried by some physical process.
Given our assumption that outside the stretched horizon, the
physics agreed more or less with known physics, it appears that
outside the stretched horizon, the only possible carrier of 
information is the Hawking radiation. This assumption does permit
that some information could be stored by unknown physics within
the stretched horizon, but relativity does not allow information within the stretched
horizon to be transmitted quickly enough for the scrambling time to be
order $M \log M$.

\section{Transmitting Information in the Photon Sphere}

We now switch to a more computer-science mode of reasoning. Suppose we have a
network of these cells, where each cell contains order 1 bits, and we wish to
send order $M^2$ bits from one side of the black hole to the other. We divide
time into time steps of size $M$, as measured by an outside observer.
Each cell can communicate with its neighbor in one time step. 

More specifically, suppose we have two hemispheres of the black hole, one near
the North pole and one near the South pole, each of which is nearly pure. Each of
these hemispheres contains half of the black
hole's suface, so each contains order $M^2$ bits. To obtain a maximally
entangled state, we need to send order $M^2$ information from one of these
regions to the other. 

\begin{figure}[tbhp]
\begin{center}
\includegraphics[width=2.4in]{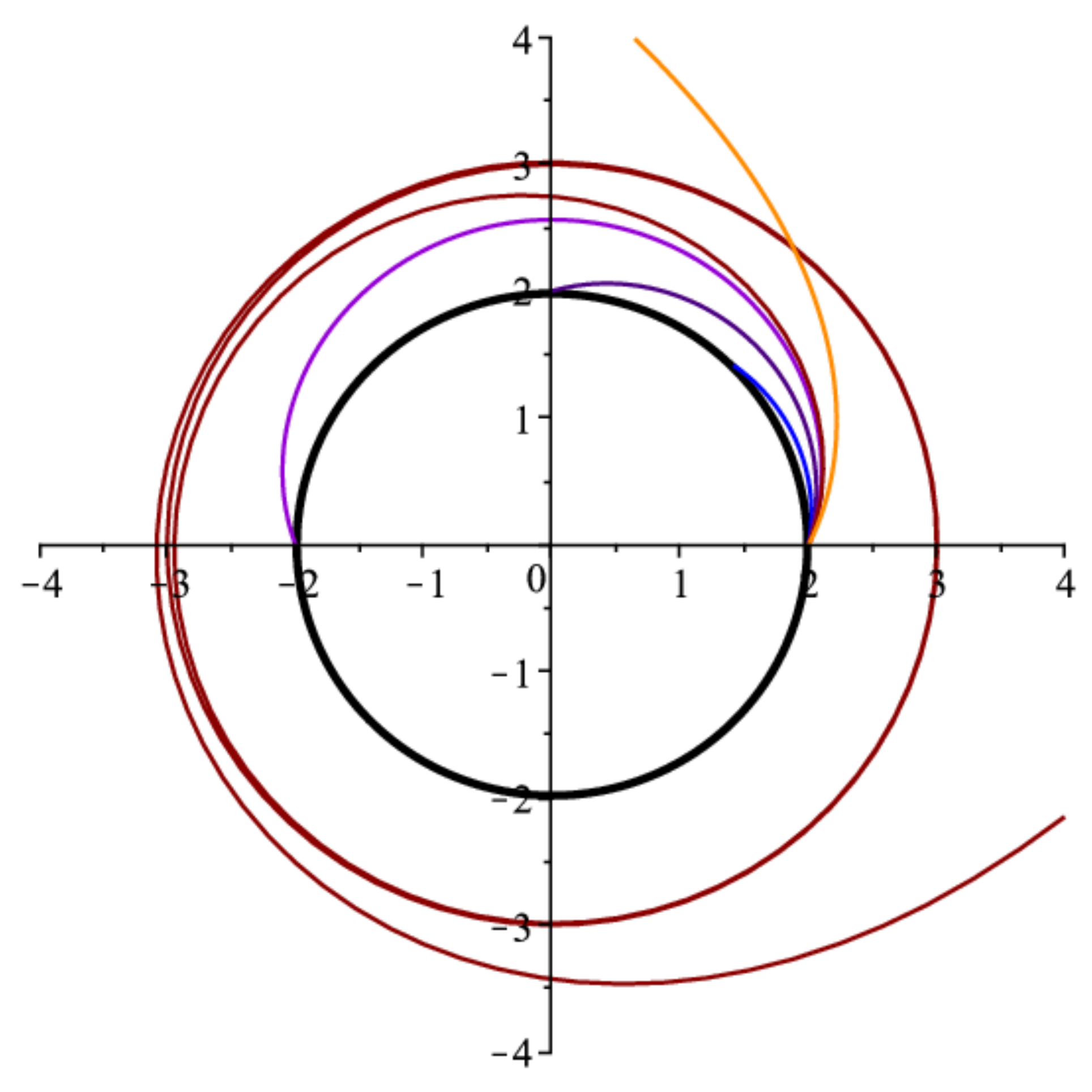}
\end{center}
\caption{Trajectories of photons from a point on the horizon of a black hole.
The red trajectory makes a complete orbit at $R=3M$ before
it escapes to infinity.
}
\label{figure-geodesics}
\end{figure}

Let us consider what paths we could send this
information along. The quickest path (as measured by an outside observer) between two points 
in the stretched horizon is the shortest null geodesic between them 
(see Figure \ref{figure-geodesics}). There is a path that might be easier to think 
about that is not much longer:  go
straight up at the speed of light until you reach a level where the two points
are covered by adjacent cells, travel at this height until you are 
directly above the second point, and then go back down.
If the two points in the stretched horizon are separated by an angle $\theta$, 
then this second path will take time order $\log(M \theta)$, and this will thus 
apply to the null geodesic as well.

So one possibility is that we send the information up to the
outer region of the photon sphere, send it along the outside of this sphere,
and then send it back down to near the horizon once it reaches the other
side. The length of these paths is  order $\log M$, which is fast enough for 
the scrambling time to be order $M \log M$.
However, there are only a
constant number of cells on the outside of the photon sphere. These cells
form a bottleneck, since each of these paths must go through one of these
cells. Each of these
cells on the outer boundary can only contain a constant number of
qubits per time step, so if we
are going to send $M^2$ qubits along paths involving these cells, 
we need to take order $M^2$ time steps, resulting in total time order
$M^3.$

Another possibility is that we send the information while keeping it close the horizon of the
black hole. There are order $M$ disjoint paths from one of these regions
to the other along the horizon, but each of these paths has length order
$M$. It would thus take order $M$ time steps 
to send all the information along these paths, making order $M^2$ time
altogether. In fact, we will show that this is the best we can
do.

Recall that we are assuming that any non-standard physics is confined to the
stretched horizon. Thus, any information that is outside the stretched horizon
must be carried by Hawking radiation. (Presumably, gravitons could also carry
information from one region of the black hole to another, but since gravitons
are massless particles, their thermodynamics should be similar to the thermodynamics
of Hawking radiation.)

Suppose we cut the black hole in half by a plane through its center. The
intersection of the cell structure with this plane will look something
like that 
in Figure 1a. There are order $M$ cells in this cut. We can see this by
observing that the inner layer has order $M$ cells, and the number of cells in
each layer forms a geometric series. If we assume that each cell can only
send a constant number of bits to its neighbors during each time step, then
each cell can
process a constant number of bits per time step. Since the time steps each
take time order $M$, and we need order $M$ of them, 
it will take order $M^2$ time to pass order $M^2$ 
bits from one side of this cut to the other. The scrambling time must be
at least this large. Thus, with our assumptions,
a lower bound for the scrambling time
is order $M^2$. Note that the same argument shows that to get $Q$ bits of
quantum information from one hemisphere of the black hole to the other,
we need time order $QM$. 

\section{Hayden and Preskill Revisited}

We have seen that to get $Q$ bits of quantum information from one hemisphere of the black
hole to the other, we need to take time order $QM$. Recall that in Hayden and Preskill's paper,
Alice tosses her diary into the black hole. Let us assume her diary weighs
$0.02$ mg---one Planck mass (rather a small mass for a diary, but we assume that
Alice writes small). How many bits does the hole receive? Ignoring the gravitational
potential energy contained in the diary, if we add $1$ Planck
mass to the hole, then the mass of the black hole increases from $M$ to $M+1$, and the
number of bits goes from $M^2$ to $M^2+2M+1$. We have thus increased the number of
bits in the black hole by $2M$. We would need to wait order $M^2$ time for all these bits to be spread
evenly through the black hole, and possibly this is the minimum amount of time it takes for
these bits to come out in the Hawking radiation.

Suppose we take one photon of light and send it into the hole, and encode a qubit
in it by arranging for it to be in some specific polarization. This doesn't improve
things much. The photon has energy $\alpha$ in Planck units. Thus, when we
add it to the black hole, the information content of the black hole increases by
$2 \alpha M$ bits, and it takes order $\alpha M^2$ time for the black hole
to scramble, which is still proportional to $M^2$, even if $\alpha$ is very small
(around $10^{-25}$ for visible light).  We do not seem to be saved by the fact
that Bob knows everything about the photon except the polarization, because
we might need to wait for all 
the extra information we've added to be evenly distributed through it, and not 
just the unknown polarization. It is possible that the one unknown qubit of polarization
scrambles
quickly despite the fact that we've added a lot of known information, but this 
would require some justification.

To add one just bit to the black hole, we do not see any better way of doing it than 
sending a photon of energy $1/(2M)$, which will have wavelength comparable to the 
radius of the black hole. The fact that we can recover a photon of wavelength
comparable to the radius in time order $M \log M$ seems much less remarkable
than recovering the polarization of a visible light photon; it seems as though
a photon of radius $2M$ 
may essentially already be spread throughout the black hole as soon as it enters it. 

\section{Discussion and Speculations}
The arguments in this paper appear to indicate that at least one of the
commonly held beliefs about black holes in the following list is incorrect:
\begin{enumerate}
	\item Black hole evolution is unitary.
	\item The causal structure in the neighborhood of a black hole is that 
	predicted by general relativity.
	\item The scrambling time of a black hole is order $M \log M$.
	\item Outside of the stretched horizon, any information in a black hole is 
	contained in the Hawking radiation. This includes the information leaving
	the black hole, and the information that scrambles it.
	\item The amount of information that Hawking radiation contains is
	the amount predicted by thermodynamics.	
\end{enumerate}
It is not clear whether this argument can be extended to narrow down the
above list of assumptions which might possibly be incorrect.

One question that could be asked is whether we can learn any more about black holes
by considering the cell structure given by this paper.
We believe that one thing the cell structure seems to indicate is that we should not think
of black hole dynamics as taking place solely on the horizon, as in the
membrane picture \cite{membranes}. In the cell structure, there are order $\log M$
layers in the photon sphere, and the outer layers seem to
play a crucial role, even though they do not contain very many
quantum bits. Without these layers it would be
difficult or impossible for an unequal distribution of charge on the surface of a black hole
to equalize quickly without a violation of
relativity, as the information about the amount of charge (or mass) added to
the black hole could not propagate from one side of the hole to the other
in time $M \log M$. With the cell structure, it is possible
to equalize the charge (or mass), as we do not
need to get much information from one side to the other; the charge and mass 
are scalar quantities, so to communicate how much charge there is,
one needs only to communicate order $\log M$ bits, which is possible in order $M
\log M$ time using the cell structure. 

Similarly, if one believes that all the information is sitting on the horizon, then in order for
the out-of-time order correlations to decay in order $M \log M$ time, either one has
to accept non-local causality, or one has to realize that information can be carried on
short paths, through the outer layers of the photon sphere (in our case, by 
virtual photons of Hawking radiation). 

Further, we believe that even near the horizon, something can be
learned about black holes
 by imagining the dynamics in three dimensions. Suppose we try to
extend the cell structure inwards beyond the Planck scale. What happens? The
cells at a Planck distance from the horizon are (as seen from an outside
observer) at Planck temperature. If we consider
possible layers closer to the
horizon, a naive application of the
formula would say that the temperature should increase. 
But what happens when you try
to add energy to a system at Planck temperature? Paradoxically, the
temperature decreases. A system at Planck temperature contains lots of
Planck-size black holes, which are constantly forming and evaporating
(or nearly forming and then evaporating, if you assume that black holes can 
never actually form in finite time). If you add energy to these, you obtain
larger-than-Planck-size black holes, which have lower temperature. These
will start absorbing mass, and increasing in size, until the ambient
temperature of the space outside them is the same as the temperature of a
black hole.  This means that in a static universe of constant volume at 
thermal equilibrium\footnote{It is
not  clear that such a thing is allowed by the laws of physics, as general
relativity may only allow static universes with the aid of a cosmological
constant, and these may be unstable}, there is
at most one black hole, and the amount of mass contained in it is a constant
fraction of the mass of the universe. This is because the black hole will
absorb radiation until the ambient temperature outside of the black hole
decreases faster than the temperature of the black hole. This can only 
happen if the black hole contains much of the mass of the universe. (If the
volume of the universe is too large compared to the mass it contains, the
black hole will evaporate completely.)

Thus, below the Planck-scale layer, we expect to find large black
holes. And indeed, it seems quite likely that below the layer of 
Planck-scale black holes just above the horizon, we indeed find the
surface of a single black hole---namely, €"the actual black 
hole.

For an eternal black hole which stays at the same mass, the black hole
horizon should be in thermodynamical equilibrium.
Thus, if it is constantly absorbing Planck-scale near-black-holes, 
one expects it to be constantly emitting them as well. (Of course, these
Planck-scale black holes may never completely form from an outside
observer's point of view.) One might thus
expect space-time at the stretched horizon, as inferred by an outside
observer, to be very irregular.
Confirmation of this and more details of this phenomenon may 
require a theory of quantum gravity.

We have identified one way in which information might be 
communicated from one side of the black hole to the order in time
less than order $M^2$. However, this is a fairly far-fetched speculation
that we believe is ruled out by several
considerations. If the Planck-scale black holes in
the atmosphere are not just black holes, but also worm-holes, then in
a mature black hole, we might expect to find wormholes connecting
one side of the black hole to the other. Information falling on one side
could then be propagated to the other side quickly. There are numerous
problems with this proposal. We believe it is very unlikely these wormholes
could last long enough to travel very far from where they are formed before
they evaporate. Further, there is a large gradient in the time dilation 
constant near the horizon; unless there is a mechanism for preventing
it \cite{KT}, space-like separated wormhole mouths would turn into time-like separated
worm-hole mouths, and information falling into them might be unavailable for
long periods of time, or might even emerge before it fell in. This would give rise
to causality violation, and it is difficult
to construct a consistent theory of physics with causality violation.
Finally, unless these wormholes last long enough that each of them can
transmit much more than a constant number of bits, arguments similar
to those in our paper show that we these wormholes cannot be moved
around the stretched horizon quickly enough to enable fast scrambling.

\section{Acknowledgements}
The author thanks Zi-Wen Liu, Seth Lloyd, Imam Marvian, Leonard Susskind, L{\'a}rus Throlacius, and
Quntao Zhuang
for helpful discussions. He
is supported by the National Science
Foundation under Grant No. CCF-1525130 and through the 
NSF Science and Technology
Center for Science of Information under Grant No.
CCF0-939370.

\end{document}